\documentclass[11pt,a4wide]{article}
\usepackage{newlfont}
\usepackage{amsmath}
\usepackage{graphics}
\usepackage{float}
\usepackage{graphicx}
\usepackage{epsfig}
\begin{document}

\title{Decay Properties of Conventional and Hybrid Charmonium Mesons}
\author{Nosheen Akbar\thanks{e mail: nosheenakbar@cuilahore.edu.pk,noshinakbar@yahoo.com} \\
\textit{Department of Physics, COMSATS University Islamabad, Lahore Campus } \\
{Defence Road, Lahore (54000), Pakistan}}
\date{}
\maketitle

\begin{abstract}
  In this paper, Schrodinger equation is numerically applied through non-relativistic potential model for deriving Spectrum, radial wave functions at origin, decay constants, lepton and photon decay widths for radial and orbital excited conventional as well as hybrid charmonium mesons. These calculated results are found in agreement with others theoretical results and with the experimental observations.
 \end{abstract}


\section*{I. Introduction}

Investigation of charmonium system (conventional and hybrid) is very important objective of particle physics. In order to study a charmonium system, spectrum, radial wave function at origin, decay constant, leptonic and photon decay are considered as important characteristics. Charmonium system consisting of bound state charm quark-antiquark pair with ground state gluonic field can be well explained using quark model\cite{gell,zweig}. In the quark model, $J^{PC}$ of quark-antiquark pair can be found by $J=L\oplus S$, $P=(-1)^{L+1}$, and $C=(-1)^{L+S}$. However, some exotic states which may be hybrid, glueballs or exotic are detected at Belle, LHC, CDF, and BESIII which remained unexplained by quark model. The charmonium system with excited state gluonic field are named as hybrid charmonium. According to literature, different models like Flux tube model \cite{flux}-\cite{flux5}, the lattice QCD \cite{LQCD}-\cite{LQCD11}, QCD string model \cite{QCDS}-\cite{QCDS1}, the quark model with a constituent gluon \cite{QM}-\cite{QM2} and the QCD sum rules \cite{QCDSR}-\cite{QCDSR5} may be used to study such hybrids. In this paper, a modified non-relativistic potential model \cite{Nosheen11, Nosheen14,Nosheen17, Nosheen19} is used to find the numerical solution of Schrodinger equation for hybrid mesons using Born Openheimer formalism and adiabatic approximation. Parity and charge of hybrid meson is found by the $ P=\varepsilon (-1)^{L+\Lambda +1}$ and  $C= \varepsilon \eta (-1)^{L+\Lambda +S}$\cite{morning}. In this formula, for the ground state gluonic field $\Lambda = 0$, while for the first excited state gluonic field, $\Lambda = 1$ and so on.  Schrodinger equation is solved numerically in order to find spectrum, radial wave function at origin, decay constant, leptonic decay width, two photon and three photon decay width of radially excited S and P states of conventional and hybrid charmonium mesons. A comparison of the results for spectrum and decay characteristics of charmonium meson $J^{PC}$ states with the experimentally known quantities with the same $J^{PC}$ may help in identifying the charmonium mesons like X(3872).

In the section II of this paper, potential models are discussed to calculate radial wave functions for the ground and radially excited state $c\overline{c}$ conventional and hybrid mesons by numerical  solution of the Schr$\ddot{o}$dinger equation. The expressions used to find radial wave function at origin, decay constant, leptonic decay, two photon decay, and three photon decay of $c\overline{c}$ mesons are written in section III, while the results are discussed in section IV.

\section*{II. Methodolgy}
Time independent Schr$\ddot{\text{o}}$dinger equation, $H \Psi = E \Psi $, can be used to find the properties of system of quark-antiquark pair. The Hamiltonean $\emph{H}$ is the energy operator and $E$ is the total energy of the system. Hamiltonian can be defined as:
\begin{equation}
H = \frac{- 1}{2 \mu}\nabla^2 + H_V +m_Q +m_{\overline{Q}}
\end{equation}
where $\mu$ is the reduced mass of the quark-antiquark system and $H_v$ is the Potential energy part of the Hamiltonian.
\subsection{Potential for Conventional $c\overline{c}$ mesons}
For conventional charmonium mesons, Potential is modelled as \cite{barnes05}:
\begin{eqnarray}
H_V =V_(r) &=&\frac{-4\alpha _{s}}{3r}+br + H^{cont}+H^{tens}+H^{s.o}.
\end{eqnarray}
Here
$\frac{-4\alpha _{s}}{3r}$ describes coulomb like interaction while linear term $b r$ is due to linear confinement. 
\begin{equation}
H^{cont} = \frac{32\pi \alpha_s}{9 m_q m_{\overline{q}}} (\frac{\sigma}{\sqrt{\pi}})^3 e^{-\sigma ^{2}r^{2}} \textbf{S}_{q}. \textbf{S}_{\overline{q}},
\end{equation}
\begin{equation}
H^{s.o} = (\frac{\textbf{S}_{q}}{4 m_q^2}  + \frac{\textbf{S}_{\overline{q}}}{4 m_{\overline{q}}^2}).\textbf{L} (\frac{4\alpha _{s}}{3r^3}- \frac{b}{r})+ \frac{\textbf{S}_{q}+ \textbf{S}_{\overline{q}}}{2 m_q m_{\overline{q}}}.{\textbf{L}}\frac{4\alpha _{s}}{3r^3}, \label{Vr}.
\end{equation}
\begin{equation}
H^{tens} = \frac{4 \alpha _{s}}{m_q m_{\overline{q}} r^3} S_T,
\end{equation}
$H^{cont}$,, $H^{s.o}$, and $H^{tens}$ describe the colour contact, spin orbit interactions, and colour tensor respectively.
$\alpha _{s}$, $b$ are the strong coupling constants and string tension while $S_T$ is the tensor operator defined as:
\begin{equation}
S_T=\textbf{S}_c.{\hat{r}}\textbf{S}_{\overline{c}}.{\hat{r}}-\frac{1}{3}\textbf{S}_{c}.
\textbf{S}_{\overline{c}},
\end{equation}
such that
\begin{equation}
<^{3}L_{J}\mid S_T\mid ^{3}L_{J}>=\Bigg \{
\begin{array}{c}
-\frac{1}{6(2L+3)},J=L+1 \\
+\frac{1}{6},J=L \\
-\frac{L+1}{6(2L-1)},J=L-1.
\end{array}
\end{equation}

\begin{equation}
\overrightarrow{L}.\overrightarrow{S}=[J(J+1)-L(L+1)-S(S+1)]/2,
\end{equation}

\begin{equation}
\overrightarrow{S}_{c}.
\overrightarrow{S}_{\overline{c}}=\frac{S(S+1)}{2}-\frac{3}{4}.
\end{equation}

Here, $L$ is the relative orbital angular momentum of the quark-antiquark and $S$ is the total spin angular momentum. Spin-ornit and colour tensor terms are equal to zero~\cite{charmonia05} for $L=0$. $m_{c}$ is the constituent mass of charm quark.

The radial Schr$\ddot{\text{o}}$dinger equation for charmonium meson can be written as:
\begin{equation}
U^{\prime \prime }(r)+2\mu (E-H_V-\frac{L(L+1)}{2\mu r^{2}})U(r)=0.
\label{P23}
\end{equation}
 Here $U(r)=rR(r)$, product of interquark distance $r$ and the radial wave function $R(r)$. At small distance(r $\rightarrow$ 0),wave function becomes unstable due to very strong attractive potential. This problem is solved by applying smearing of position co-ordinates by using the method discussed in ref. \cite{godfrey}. The parameters ($ m_c, \alpha_s, b,\sigma$) are found by fitting the meson's mass with experimentally known mass, we got the following values. $ m_c = 1.454 GeV, \alpha_s= 0.5315$, $\sigma=1.105 GeV$,and $b=0.1583$ $\text{GeV}^2$. With these parameters, masses of $S$ and $P$ states are calculated which are very close to the predictions of other theoretical works as shown in Table 1.

\subsection{Potential Model for $c\overline{c}$ Hybrid Meson}
 The extended form of conventional meson potential model\cite{Nosheen11} is used to study the hybrid meson. In this extended model, an additional term ($\frac{c}{r} + A\times \exp^{-Br^{0.3723}}$) is added in the conventional meson potential model. Values of parameters $A = 3.4693 GeV$, $B  =1.0110 GeV$, and $c = 0.1745$ are taken from our earlier fit\cite{Nosheen11} to the lattice data \cite{morning} of the parameters of the
effective potential form corresponding to the first excited gluonic state. For hybrid mesons, radial Schrodinger equation is written as:
\begin{equation}
U^{\prime \prime }(r)+2\mu \left( E-V(r)-\frac{c}{r}-A\times \exp^{-Br^{0.3723}}-\frac{L(L+1)-2\Lambda
^{2}+\left\langle J_{g}^{2}\right\rangle}{2\mu r^{2}}\right) U(r)=0,
\end{equation}
For the first gluonic excitation, the square of gluon angular momentum $\left\langle J_{g}^{2}\right\rangle =2$ and projection of gluon angular momentum $\Lambda =1$~\cite{morning} making $-2\Lambda ^{2}+\langle J_{g}^{2}\rangle =0 $.
The above equation is solved by shooting method to find the spectrum of hybrid $c\overline{c}$ meson. Mass of hybrid mesons are reported in Table 2.

\begin{table}\caption{Masses of ground, radially, and orbitally excited state $c\overline{c}$ mesons . }
\tabcolsep=4pt
\fontsize{8}{10}\selectfont
\begin{center}
\begin{tabular}{|c|c|c|c|c|c|} \hline
 Meson &  $J^{PC}$ & Calculated mass & Theor. mass~\cite{charmonia05} & \cite{Nosheen14}& Exp. mass \\
 & &  \textrm{GeV} &\textrm{GeV} & \textrm{GeV} & \textrm{GeV}\\ \hline
$ \eta_{c} (1 ^1S_0)$ & $0^{-+}$ &2.9808 & 2.982 & 2.9816 &$2.9839\pm 0.0005$~\cite{pdg} \\
$J/\psi (1 ^3S_1)$& $1^{--}$ & 3.0903& $3.090$ & 3.0900 & $3.0969\pm 0.000006$~\cite{pdg} \\ \hline
$ \eta'_{c} (2 ^1S_0)$ & $0^{-+}$ & 3.656&3.630 & 3.6303 & $3.6375\pm 0.0011 $~\cite{pdg} \\
 $J/\psi (2 ^3S_1)$& $ 1^{--}$ & 3.6997 & 3.672 & 3.6718 & $3.6861 \pm 0.000025 $ ~\cite{pdg}\\ \hline
$ \eta_{c} (3 ^1S_0)$ &$0^{-+}$ & 4.0955& 4.043 & 4.0432 &--- \\
$J/\psi (3 ^3S_1)$& $1^{--}$ & 4.0709& 4.072 & 4.0716 & $4.040 \pm 10$~\cite{charmonia05}\\ \hline
$ \eta_{c} (4 ^1S_0)$ & $0^{-+}$ & 4.4599& 4.384 & 4.3837 &--- \\
$J/\psi (4 ^3S_1)$& $1^{--}$ & 4.4838 & 4.406 & 4.4061 &$4.415 \pm 6$~\cite{charmonia05} \\ \hline
$ \eta_{c} (5 ^1S_0)$ &0$^{-+}$ & 4.7831& --- & 4.6850 &--- \\
$J/\psi (5 ^3S_1)$& $1^{--}$ & 4.8032 & --- & 4.7038 & \\ \hline
$ \eta_{c} (6 ^1S_0)$ &0$^{-+}$ &  5.079 &--- & 4.9604 &--- \\
$J/\psi (6 ^3S_1)$& $1^{--}$ & 5.0966 & --- & 4.9769 & ---\\ \hline
$ h_{c} (1 ^1P_1) $&$1^{+-}$ & 3.5288& $3.516$ & 3.5156 & $3.52538\pm 0.00011$~\cite{pdg}\\
$\chi_{0} (1 ^3P_0)$ & $0^{++}$ & 3.4526& $3.424$ & 3.4245 & $3.4147\pm0.00030$ ~\cite{pdg}\\
$\chi_{1} (1 ^3P_1)$ &$1^{++}$ & 3.5244 & $3.505$ & 3.5054 & $3.51067 \pm 0.00007$~\cite{pdg} \\
$\chi_{2} (1 ^3P_2)$ &$2^{++}$ & 3.5626& $3.556$ & 3.5490 & $3.55617 \pm 0.00007$~\cite{pdg} \\ \hline
$h_{c} (2 ^1P_1) $& $1^{+-}$& 3.9747 & 3.934 & 3.9336 &--- \\
$\chi_{0} (2 ^3P_0)$ &$ 0^{++}$ & 3.9453 & 3.852 & 3.8523 &--- \\
$\chi_{1} (2 ^3P_1)$ &$1^{++}$ &  3.980 & 3.925 & 3.9249 &---\\
$\chi_{2} (2 ^3P_2)$ &$2^{++}$ &  3.9957 & 3.972 & 3.9648 & $3.9272 \pm 0.0026$~\cite{pdg} \\ \hline
$h_{c} (3 ^1P_1) $&$1^{+-}$ & 4.3452 &4.279 & 4.2793 & --- \\
$\chi_{0} (3 ^3P_0)$ &$0^{++}$ & 4.3296& 4.202 & 4.2017 &---\\
$\chi_{1} (3 ^3P_1)$ &$1^{++}$ & 4.3532 & 4.271 & 4.2707 &---\\
$\chi_{2} (3 ^3P_2)$ &$2^{++}$ & 4.3619& 4.317 & 4.3093  & ---\\ \hline
$h_{c} (4 ^1P_1) $& $1^{+-}$ & 4.6735 & --- &4.5851 & --- \\
$\chi_{0} (4 ^3P_0)$ &$0^{++}$ & 4.664 & & 4.5092 &---\\
$\chi_{1} (4 ^3P_1)$ &$1^{++}$ &4.6823& --- &4.5762 & ---\\
$\chi_{2} (4 ^3P_2)$ &$ 2^{++}$& 4.688& --- &4.6141 & ---\\ \hline
$h_{c} (5 ^1P_1) $&$1^{+-}$ & 4.9737& --- & 4.8644 &---\\
$\chi_{0} (5 ^3P_0)$ &$0^{++}$ & 4.9675 & --- & 4.7894 &---\\
$\chi_{1} (5 ^3P_1)$ &$1^{++}$ & 4.9827 & --- & 4.8552 &---\\
$\chi_{2} (5 ^3P_2)$ &$ 2^{++}$& 4.9868 & --- & 5.8926 &---\\ \hline
$h_{c} (6 ^1P_1) $&$1^{+-}$ & 5.2534& --- & 5.1244 &--- \\
$\chi_{0} (6 ^3P_0)$ &$0^{++}$ & 5.2493& --- & 5.0500 &---\\
$\chi_{1} (6 ^3P_1)$ &$1^{++}$ & 5.2623& --- & 5.1148 &---\\
$\chi_{2} (6 ^3P_2)$ & $ 2^{++}$& 5.2654 & --- & 5.1520 &---\\ \hline
\end{tabular}
\end{center}
\end{table}

\begin{table}\caption{Mass of $c\overline{c}$ hybrid mesons. }
\begin{center}
\begin{tabular}{|c|c|c|c|c|c|c|}
\hline
 Meson & \multicolumn{2}{|c|}{$J^{PC}$}& Calculated Mass & \cite{Nosheen14} & \cite{Isgur1} & \cite{LIU}\\
 & $\varepsilon=1$& $\varepsilon=-1$ &  & & &  \\
 & & &  \textrm{GeV}&  \textrm{GeV} & \textrm{GeV} & \textrm{GeV}\\ \hline
$ \eta^{h}_{c} (1 ^1S_0)$ &$0^{++}$ & $0^{--}$ &4.0866 &4.0802 & & \\
$ J/\psi^{h} (1 ^3S_1)$& $1^{+-}$ & $1^{-+}$ &4.1158 & 4.1063 & 4.19& 4.213 \\ \hline
 $ \eta^{h}_{c} (2 ^1S_0)$ &$0^{++}$ & $0^{--}$ &4.4198& 4.3820 & & \\
 $ J/\psi^{h} (2 ^3S_1)$& $ 1^{+-}$ & $1^{-+}$ & 4.4484& 4.4084 & & \\ \hline
 $ \eta^{h}_{c} (3 ^1S_0)$ &$0^{++}$ & $0^{--}$ & 4.7254& 4.6616 & &\\
 $ J/\psi^{h} (3 ^3S_1)$& $1^{+-}$ & $1^{-+}$ & 4.751 & 4.6855& & \\ \hline
 $ \eta^{h}_{c} (4 ^1S_0)$ &$0^{++}$ & $0^{--}$ & 5.0094 & 4.9223 & &\\
 $ J/\psi^{h} (4 ^3S_1)$& $1^{+-}$ & $1^{-+}$ &5.0323 & 4.9438& & \\ \hline
 $ \eta^{h}_{c} (5 ^1S_0)$&$0^{++}$ & $0^{--}$ & 5.2766 & 4.1683& &\\
 $ J/\psi^{h} (5 ^3S_1)$& $1^{+-}$ & $1^{-+}$ & 5.2972& 4.1876& & \\ \hline
 $ \eta^{h}_{c} (6 ^1S_0)$ &$0^{++}$ & $0^{--}$ &5.5303 & 4.4021 & &\\
 $ J/\psi^{h} (6 ^3S_1)$& $1^{+-}$ & $1^{-+}$ & 5.549& 4.4197& & \\ \hline
 $ h^{h}_{c} (1 ^1P_1) $& $1^{--}$ & $1^{++}$ & 4.2943 & 4.2678& 4.19 &\\
 $ \chi^{h}_{0} (1 ^3P_0)$& $0^{-+}$ & $0^{+-}$ &4.2447 & 4.2464  & 4.19 &4.382 \\
 $ \chi^{h}_{1} (1 ^3P_1)$ &$1^{-+}$ & $1^{+-}$ & 4.2893 & 4.2678& & \\
 $ \chi^{h}_{2} (1 ^3P_2)$ &$2^{-+}$ & $2^{+-}$ & 4.3093 & 4.2739& 4.19& 4.391\\ \hline
 $ h^{h}_{c} (2 ^1P_1) $&  $1^{--}$& $1^{++}$ & 4.6085 & 4.5552& & \\
 $ \chi^{h}_{0} (2 ^3P_0)$ &$ 0^{-+}$ & $0^{+-}$ &4.5812 & 4.5264 & &\\
 $ \chi^{h}_{1} (2 ^3P_1)$ &$1^{-+}$ & $1^{+-}$ &4.6086& 4.5538 & &\\
 $ \chi^{h}_{2} (2 ^3P_2)$ &$2^{-+}$ & $2^{+-}$ &4.6201& 4.5653& &4.505  \\ \hline
 $ h^{h}_{c} (3 ^1P_1) $&  $1^{--}$ & $1^{++}$ &4.8981& 4.8210 & & \\
 $ \chi^{h}_{0} (3 ^3P_0)$ & $0^{-+}$ & $0^{+-}$ & 4.8797 & 4.7875& &\\
 $ \chi^{h}_{1} (3 ^3P_1)$ &$1^{-+}$ & $1^{+-}$ & 4.9006& 4.8188 & &\\
 $ \chi^{h}_{2} (3 ^3P_2)$ &$2^{-+}$ & $2^{+-}$ & 4.9085 & 4.8337& &\\ \hline
 $ h^{h}_{c} (4 ^1P_1) $ & $1^{--}$ & $1^{++}$ &5.1695& 5.0707  & &\\
 $ \chi^{h}_{0} (4 ^3P_0)$ &$0^{-+}$ & $0^{+-}$ & 5.1562 & 5.0338& &\\
 $ \chi^{h}_{1} (4 ^3P_1)$  &$1^{-+}$ & $1^{+-}$ & 5.1735 & 5.0678& &\\
 $ \chi^{h}_{2} (4 ^3P_2)$ &$ 2^{-+}$& $2^{+-}$ & 5.1793& 5.0852& &\\ \hline
 $ h^{h}_{c} (5 ^1P_1) $&  $1^{1--}$ & $1^{++}$ & 5.4266 & 5.3076 & &\\
 $ \chi^{h}_{0} (5 ^3P_0)$ &$0^{-+}$ & $0^{+-}$ & 5.4168 & 5.2682 & &\\
 $ \chi^{h}_{1} (5 ^3P_1)$ &$1^{-+}$ & $1^{+-}$ & 5.4316 & 5.3042& &\\
 $ \chi^{h}_{2} (5 ^3P_2)$ &$ 2^{-+}$& $2^{+-}$ & 5.4361 & 5.3233& &\\ \hline
 $ h^{h}_{c} (6 ^1P_1) $&  $1^{--}$ & $1^{++}$ & 5.6722& 5.5340 & &\\
 $ \chi^{h}_{0} (6 ^3P_0)$ &$0^{-+}$ & $0^{+-}$ & 5.6648 & 5.4925& &\\
 $ \chi^{h}_{1} (6 ^3P_1)$ &$1^{-+}$ & $1^{+-}$ & 5.6778& 5.5301 & &\\
 $ \chi^{h}_{2} (6 ^3P_2)$ &$ 2^{-+}$& $2^{+-}$ & 5.6814 & 5.5507& &\\ \hline
\end{tabular}
\end{center}
\end{table}

\begin{table}\caption{S state of radial wave function at origin and decay constant of $c\overline{c}$ mesons . }
\tabcolsep=4pt
\fontsize{9}{11}\selectfont
\begin{center}
\begin{tabular}{|c|c|c|c|c|c|c|c|}
\hline
Meson & $J^{P}$& Calculated & $|R(0)|^2$ & Calculated &$f_p$ &$f_p$ &$f_p$ \\
State & &  $|R(0)|^2$ & \cite{Nosheen14} & $f_p$  & \cite{EPJ592} &\cite{decay} &\cite{PDG} \\ \hline
 & & $\textrm{GeV}^3$&$\textrm{GeV}^3$ &$\textrm{GeV}$ & $\textrm{GeV}$ & $\textrm{GeV}$ &  $\textrm{GeV}$\\ \hline
 $ (1 ^3S_1)$ & $1^{--}$ & 1.1886 & 1.2294 &0.3326 &0.325 & & $0.416 \pm 0.006$\\
$ (1 ^1S_0)$ & $0^{-+}$ &1.9125 & 1.9768 &0.519 &  0.350 &0.387 & $0.335\pm 0.075$ \\ \hline
 $(2 ^3S_1)$& $1^{--}$ &0.7287 & 0.7225& 0.238 &0.257 & &$0.304\pm 0.304$ \\
$ (2 ^1S_0)$ & $0^{-+}$ &0.8850 & 0.8717&0.3181 & 0.278& &\\ \hline
 $(3 ^3S_1)$& $1^{--}$ & 0.6136& 0.6006& 0.2069 & 0.229& &\\
 $(3 ^1S_0)$ & $0^{-+}$ &0.7011 &0.683 &0.2675 & 0.249 & &\\ \hline
 $(4 ^3S_1)$& $1^{--}$ & 0.6186& 0.5994& 0.189 & 0.212& & \\
 $(4 ^1S_0)$ & $0^{-+}$ & 0.5571& 0.5417& 0.2408& 0.231& &\\ \hline
 $(5 ^3S_1)$& $1^{--}$ & 0.5220& 0.5503& 0.1768& 0.200& & \\
 $(5 ^1S_0)$ & $0^{-+}$ &0.5699 & 0.5417& 0.2232& 0.218 & &\\ \hline
 $(6 ^3S_1)$& $1^{--}$ & 0.4975& 0.5172-& 0.1676&0.191 & &\\
 $(6 ^1S_0)$ & $0^{-+}$ & 0.5371& 0.5053& 0.2102& 0.208 & &\\ \hline
\end{tabular}
\end{center}
\end{table}

\begin{table}\caption{Radial wave function at origin and decay constant for S state of $c\overline{c}$ hybrid mesons. }
\begin{center}
\begin{tabular}{|c|c|c|c|c|c|}
\hline
 Meson & \multicolumn{2}{|c|}{$J^{PC}$}& Calculated $|R(0)|^2$& $|R(0)|^2$ \cite{Nosheen14} & Calculated $f_p$\\
 & $\varepsilon=1$& $\varepsilon=-1$ &  &  &\\
 & & &  \textrm{GeV}&\textrm{GeV} & \textrm{GeV}\\ \hline
$ \eta^{h}_{c} (1 ^1S_0)$ &$0^{++}$ & $0^{--}$ & 0.3294& 0.3046 & 0.1836  \\
$ J/\psi^{h} (1 ^3S_1)$& $1^{+-}$ & $1^{-+}$ &0.1687 & 0.1533 & 0.1086 \\ \hline
 $ \eta^{h}_{c} (2 ^1S_0)$ &$0^{++}$ & $0^{--}$ &0.3501 & 0.3306 & 0.182\\
 $ J/\psi^{h} (2 ^3S_1)$& $ 1^{+-}$ & $1^{-+}$ & 0.2153 & 0.1995 & 0.118\\ \hline
 $ \eta^{h}_{c} (3 ^1S_0)$ &$0^{++}$ & $0^{--}$ & 0.3442 &0.3295 & 0.1745 \\
 $ J/\psi^{h} (3 ^3S_1)$& $1^{+-}$ & $1^{-+}$ & 0.2376 & 0.2214 & 0.1199\\ \hline
 $ \eta^{h}_{c} (4 ^1S_0)$ &$0^{++}$ & $0^{--}$ & 0.3368 & 0.3189 & 0.1676  \\
 $ J/\psi^{h} (4 ^3S_1)$& $1^{+-}$ & $1^{-+}$ &0.2507 & 0.2342 & 0.1197 \\ \hline
 $ \eta^{h}_{c} (5 ^1S_0)$&$0^{++}$ & $0^{--}$ & 0.3305 & 0.312 & 0.1618\\
 $ J/\psi^{h} (5 ^3S_1)$& $1^{+-}$ & $1^{-+}$ & 0.2592 & 0.2425 & 0.1186 \\ \hline
 $ \eta^{h}_{c} (6 ^1S_0)$ &$0^{++}$ & $0^{--}$ &0.3254 & 0.3078 & 0.1568\\
 $ J/\psi^{h} (6 ^3S_1)$& $1^{+-}$ & $1^{-+}$ & 0.2650& 0.2482 & 0.1298 \\ \hline
 \end{tabular}
\end{center}
\end{table}

\begin{table}\caption{$|R'(0)|^2$ and decay constant of $c\overline{c}$ mesons . }
\tabcolsep=4pt
\fontsize{9}{11}\selectfont
\begin{center}
\begin{tabular}{|c|c|c|c|}
\hline
Meson & $J^{P}$& Calculated $|R'(0)|^2$ & Calculated $f_{\chi}$ \\
& &  $\textrm{GeV}^5$ & \textrm{GeV} \\ \hline
 $ h_{c} (1 ^1P_1) $ & $1^{+-}$ & 0.4551 & \\
 $\chi_{0} (1 ^3P_0)$ & $0^{++}$ & 0.4074 &0.7547 \\
 $\chi_{1} (1 ^3P_1)$ & $1^{++}$ & 0.3334 & 0.7883  \\
 $\chi_{2} (1 ^3P_2)$ & $2^{++}$ & 0.3021 & \\ \hline
 $h_{c} (2 ^1P_1) $ & $1^{+-}$& 0.6116 & \\
 $\chi_{0} (2 ^3P_0)$ &$ 0^{++}$ & 0.5066 & 0.8415 \\
 $\chi_{1} (2 ^3P_1)$ &$1^{++}$ &  0.4474 &0.9132 \\
 $\chi_{2} (2 ^3P_2)$ &$2^{++}$ &  0.4316 & \\ \hline
 $h_{c} (3 ^1P_1) $&  $1^{+-}$ & 0.716 & \\
 $\chi_{0} (3 ^3P_0)$ & $0^{++}$ & 0.5789& 0.8996 \\
 $\chi_{1} (3 ^3P_1)$ &$1^{++}$ & 0.531& 0.9949 \\
 $\chi_{2} (3 ^3P_2)$ &$2^{++}$ & 0.5244 & \\ \hline
 $h_{c} (4 ^1P_1) $&  $1^{+-}$ & 0.7986  & \\
 $\chi_{0} (4 ^3P_0)$ &$0^{++}$ & 0.6444 & 0.9491 \\
 $\chi_{1} (4 ^3P_1)$  &$1^{++}$ &0.6021& 1.0594 \\
 $\chi_{2} (4 ^3P_2)$ &$ 2^{++}$& 0.6011& --- \\ \hline
 $h_{c} (5 ^1P_1) $&  $1^{+-}$ & 0.869& --- \\
 $\chi_{0} (5 ^3P_0)$ &$0^{++}$ & 0.7041& 0.9921 \\
 $\chi_{1} (5 ^3P_1)$ &$1^{++}$ & 0.6654 & 1.1137 \\
 $\chi_{2} (5 ^3P_2)$ &$ 2^{++}$& 0.6687 & --- \\ \hline
 $h_{c} (6 ^1P_1) $&  $1^{+-}$ & 0.9315& --- \\
 $\chi_{0} (6 ^3P_0)$ &$0^{++}$ & 0.7593& 1.0303 \\
 $\chi_{1} (6 ^3P_1)$ &$1^{++}$ & 0.7234& 1.1612 \\
 $\chi_{2} (6 ^3P_2)$ & $ 2^{++}$& 0.7299 & --- \\ \hline
\end{tabular}
\end{center}
\end{table}

\subsection{Radial wave function at origin}
For normalized wave function:
\begin{equation}
U^{'}(0)= R(0)= \sqrt{4 \pi} \psi(0).
\end{equation}
 $U^{'}(0)$ is calculated to find the radial wave function at origin whose magnitudes for conventional and hybrid meson are reported in Table (3,4). For the states with $L>0$, wave function becomes zero at the origin. However derivatives of radial wave functions are non-zero as reported in Table (5,6).
\section*{III. Decay Properties}

\subsection{Decay Constants}
 Decay constant is an important characteristic of mesons. Decay constants ($f_p$) of pseudo scalar and pseudo vector mesons depend on $|R(0)|^2 $. Following Van-Royen-Weisskopf formula \cite{VWF} is used to find decay constants.
\begin{equation}
f_p = \sqrt{\frac{3 |R(0)^2|}{\pi M_p}} =\sqrt{\frac{12 |\psi(0)^2|}{M_p}}.
\end{equation}
where $M_p$ is the mass of corresponding meson. To calculate decay constant, numerically calculated mass (in Table 1 and 2) is used. By incorporating the first order QCD correction factor, the decay constant can be written as:
\begin{equation}
f_p = \sqrt{\frac{3 |R(0)^2|}{\pi M_p}} =\sqrt{\frac{12 |\psi(0)^2|}{M_p}} (1-\triangle \frac{\alpha_s}{\pi}),
\end{equation}
where $\triangle=2$ for $^1 S_0$ mesons and $\triangle=8/3$ for $^3 S_1$ mesons.  For P state charmonium mesons, decay constant depends on the derivative of radial wave function at origin. Following relations are used to find decay constants of $\chi_0$ and $\chi_1$ \cite{lans}
\begin{equation}
f_{\chi_0} = \sqrt{\frac{27 |R'(0)^2|}{2 \pi m^3_Q}}
\end{equation}

\begin{equation}
f_{\chi_1} = \sqrt{\frac{18 |R'(0)^2|}{\pi m^3_Q}}
\end{equation}
Decay constants for conventional and hybrid mesons are reported in Tables(3-6). Our results are very close to experimental and theoretical results.
\subsection{Leptonic Decay}
Leptonic decay width of $^3S_1$ state of charmonium meson with $J^{PC}=1^{--}$ can be calculated by the following relation defined in ~\cite{VWF,EPJ592}
\begin{equation}
\Gamma_{ee}(n ^3S_1) = \frac{4 \alpha^2 e^2_c} {M^2_ns} \mid R_{nS}(0)\mid^2 (1 - \frac{16}{3} \frac{\alpha_S}{\pi})
\end{equation}
Here, $\alpha=\frac{1}{137}$, $e_c=\frac{2}{3}$ is the charge of $c$ quark.
\subsection{Two photon Decay}
Two photon decay for S and P states is proportional to $\alpha^2$ and can be calculated by using following expression ~\cite{EPJ592,VWF}:
\begin{equation}
\Gamma(n ^1 S_0 \rightarrow \gamma \gamma) = \frac{3 \alpha^2 e^4_c \mid R_{nS}(0)\mid^2}{m^2_c} (1-\frac{3.4 \alpha_s}{\pi}),
\end{equation}
\begin{equation}
\Gamma(n ^3 P_0 \rightarrow \gamma \gamma) = \frac{27 \alpha^2 e^4_c \mid R'_{nP}(0)\mid^2}{m^4_c}(1+\frac{0.2 \alpha_s}{\pi}),
\end{equation}
\begin{equation}
\Gamma(n ^3 P_2 \rightarrow \gamma \gamma) = \frac{36 \alpha^2 e^4_c \mid R'_{nP}(0)\mid^2}{5 m^4_c} (1-\frac{16 \alpha_s}{3 \pi}),
\end{equation}

\subsection{Three photon Decay}

The decays $^3 S_1 \rightarrow \gamma \gamma \gamma $ have very small rates proportional to $\alpha^3$. In ref.\cite{3foton}, three photon decay is written as:
\begin{equation}
\Gamma(^3 S_1 \rightarrow \gamma \gamma \gamma)= \frac{4 (\pi^2 -9) \alpha^3 e^6_c}{3 \pi m_c^2} \mid R_{nS}(0)\mid^2
\end{equation}

\begin{table}\caption{Radial wave function at origin and decay constant for P state of $c\overline{c}$ hybrid mesons. }
\begin{center}
\begin{tabular}{|c|c|c|c|c|}
\hline
 Meson & \multicolumn{2}{|c|}{$J^{PC}$}& Calculated $|R'(0)|^2 $ & Calculated $f_{\chi}$\\
 & $\varepsilon=1$& $\varepsilon=-1$ &  &  \\
 & & &  \textrm{GeV}&  \textrm{GeV} \\ \hline
 $ h^{h}_{c} (1 ^1P_1) $& $1^{--}$ & $1^{++}$ & 0.055 &  \\
 $ \chi^{h}_{0} (1 ^3P_0)$& $0^{-+}$ & $0^{+-}$ &0.0634 & 0.2977 \\
 $ \chi^{h}_{1} (1 ^3P_1)$ &$1^{-+}$ & $1^{+-}$ & 0.0438 & 0.2857\\
 $ \chi^{h}_{2} (1 ^3P_2)$ &$2^{-+}$ & $2^{+-}$ & 0.035 & \\ \hline
 $ h^{h}_{c} (2 ^1P_1) $&  $1^{--}$& $1^{++}$ & 0.1163 & \\
 $ \chi^{h}_{0} (2 ^3P_0)$ &$ 0^{-+}$ & $0^{+-}$ &0.1087 &0.3898  \\
 $ \chi^{h}_{1} (2 ^3P_1)$ &$1^{-+}$ & $1^{+-}$ & 0.0876 & 0.4041 \\
 $ \chi^{h}_{2} (2 ^3P_2)$ &$2^{-+}$ & $2^{+-}$ &0.0787&  \\ \hline
 $ h^{h}_{c} (3 ^1P_1) $&  $1^{--}$ & $1^{++}$ &0.1764&  \\
 $ \chi^{h}_{0} (3 ^3P_0)$ & $0^{-+}$ & $0^{+-}$ & 0.1531 &0.4626 \\
 $ \chi^{h}_{1} (3 ^3P_1)$ &$1^{-+}$ & $1^{+-}$ & 0.1309& 0.494\\
 $ \chi^{h}_{2} (3 ^3P_2)$ &$2^{-+}$ & $2^{+-}$ & 0.123 & \\ \hline
 $ h^{h}_{c} (4 ^1P_1) $ & $1^{--}$ & $1^{++}$ &0.2335&   \\
 $ \chi^{h}_{0} (4 ^3P_0)$ &$0^{-+}$ & $0^{+-}$ & 0.1957 & 0.523\\
 $ \chi^{h}_{1} (4 ^3P_1)$  &$1^{-+}$ & $1^{+-}$ & 0.1730 & 0.5678\\
 $ \chi^{h}_{2} (4 ^3P_2)$ &$ 2^{-+}$& $2^{+-}$ & 0.1665 &  \\ \hline
 $ h^{h}_{c} (5 ^1P_1) $&  $1^{1--}$ & $1^{++}$ & 0.2874 &  \\
 $ \chi^{h}_{0} (5 ^3P_0)$ &$0^{-+}$ & $0^{+-}$ & 0.2367 & 0.5752 \\
 $ \chi^{h}_{1} (5 ^3P_1)$ &$1^{-+}$ & $1^{+-}$ & 0.2138 & 0.6313\\
 $ \chi^{h}_{2} (5 ^3P_2)$ &$ 2^{-+}$& $2^{+-}$ & 0.2089 & \\ \hline
 $ h^{h}_{c} (6 ^1P_1) $&  $1^{--}$ & $1^{++}$ & 0.3383 &  \\
 $ \chi^{h}_{0} (6 ^3P_0)$ &$0^{-+}$ & $0^{+-}$ & 0.2764 & 0.6216\\
 $ \chi^{h}_{1} (6 ^3P_1)$ &$1^{-+}$ & $1^{+-}$ & 0.2535& 0.6874 \\
 $ \chi^{h}_{2} (6 ^3P_2)$ &$ 2^{-+}$& $2^{+-}$ & 0.2500 & \\ \hline
 \end{tabular}
\end{center}
\end{table}

\begin{table}\caption{Leptonic decay width for $^3 S_1$ state conventional $c\overline{c}$ mesons.}
\begin{center}
\begin{tabular}{|c|c|c|c|c|} \hline
Decay Modes &  Calculated $\Gamma_{ee}$ & Others \cite{EPJ592} & Others\cite{patel}  & Experimental\cite{PDG} \\ \hline

& $keV$ & $keV$&$keV$ &$keV$  \\ \hline
$1^3S_1 \rightarrow e^+ e^-$ & 1.8532 &2.925 & 6.99 &$5.547\pm 0.14$ \\
$2 ^3S_1 \rightarrow e^+ e^-$ & 0.5983 & 1.533&2.38 &$2.359 \pm 0.04$ \\
$3 ^3S_1 \rightarrow e^+ e^-$ & 0.3812 & 1.091& 2.31& $0.86 \pm 0.07$ \\
$4 ^3S_1 \rightarrow e^+ e^-$ & 0.2847 & 0.856& 1.78& $0.58 \pm 0.07$ \\
$5 ^3S_1 \rightarrow e^+ e^-$ & 0.2286 & 0.707&1.46 & \\
$6 ^3S_1 \rightarrow e^+ e^-$ & 0.1913 & 0.602& 1.24 & \\ \hline
 \end{tabular}
\end{center}
\end{table}

\begin{table}\caption{Leptonic decay width for hybrid $c\overline{c}$ mesons. }
\begin{center}
\begin{tabular}{|c|c|} \hline
Decay Modes &  Calculated $\Gamma_{ee}$ \\ \hline
& $keV$ \\ \hline
$1 ^{h3}S_1 \rightarrow \gamma \gamma$ & 0.0922 \\
$2 ^{h3}S_1 \rightarrow \gamma \gamma$ & 0.1007 \\
$3 ^{h3}S_1 \rightarrow \gamma \gamma$ & 0.0974 \\
$4 ^{h3}S_1 \rightarrow \gamma \gamma$ & 0.0916 \\
$5 ^{h3}S_1 \rightarrow \gamma \gamma$ & 0.0855 \\
$6 ^{h3}S_1 \rightarrow \gamma \gamma$ & 0.0796 \\ \hline
 \end{tabular}
\end{center}
\end{table}

\begin{table}\caption{Two photon decay for $\eta_c$ state conventional and hybrid $c\overline{c}$ mesons in $keV$. }
\begin{center}
\begin{tabular}{|c|c|c|c|c|c|c|}
Decay Modes &  Calculated $\Gamma_{\gamma\gamma}$ & \cite{EPJ592} & 0thers\cite{Lakhina} & Experimental\cite{PDG} & Decay Mode & Calculated $\Gamma_{\gamma\gamma}$ for hybrid \\ \hline
$1^1S_0 \rightarrow \gamma \gamma$ & 12.1326 & 7.231& 7.18 &$5.1\pm 0.4$ & $1 ^{h1}S_0 \rightarrow \gamma \gamma$ & 2.0897\\
$2 ^1S_0 \rightarrow \gamma \gamma$ & 5.6143 & 5.501 & 1.71 & $2.15\pm 1.58$ &$2 ^{h1}S_0 \rightarrow \gamma \gamma$ & 2.221\\
$3 ^1S_0 \rightarrow \gamma \gamma$ & 4.4477 & 4.971& 1.21 & & $3 ^{h1}S_0 \rightarrow \gamma \gamma$ & 2.1836 \\
$4 ^1S_0 \rightarrow \gamma \gamma$ & 3.9243 & 4.688 & & &$4 ^{h1}S_0 \rightarrow \gamma \gamma$ & 2.1366 \\
$5 ^1S_0 \rightarrow \gamma \gamma$ & 3.6154 & 4.507 & & &$5 ^{h1}S_0 \rightarrow \gamma \gamma$ & 2.0966\\
$6 ^1S_0 \rightarrow \gamma \gamma$ & 3.4073 & 4.377 & & &$6 ^{h1}S_0 \rightarrow \gamma \gamma$ & 2.0643\\ \hline \hline
$1 ^3P_0 \rightarrow \gamma \gamma$ & 26.7777 & 8.982&3.28 &$2.34 \pm 0.19$ & $1 ^{h3}P_0 \rightarrow \gamma \gamma$ & 4.1672\\
$2 ^3P_0 \rightarrow \gamma \gamma$ & 33.2979 & 9.111& & &$2 ^{h3}P_0 \rightarrow \gamma \gamma$ & 7.1447\\
$3 ^3P_0 \rightarrow \gamma \gamma$ & 38.0501 & 9.104& & &$3 ^{h3}P_0 \rightarrow \gamma \gamma$ & 10.063\\
$4 ^3P_0 \rightarrow \gamma \gamma$ & 42.3553 &9.076 & & &$4 ^{h3}P_0 \rightarrow \gamma \gamma$ & 12.863\\
$5 ^3P_0 \rightarrow \gamma \gamma$ & 46.2792 &9.047 & & &$5 ^{h3}P_0 \rightarrow \gamma \gamma$ & 15.5579\\
$6 ^3P_0 \rightarrow \gamma \gamma$ & 61.2258 & & & &$6 ^{h3}P_0 \rightarrow \gamma \gamma$ & 18.1673 \\ \hline \hline
$1 ^3P_2 \rightarrow \gamma \gamma$ & 0.5004 &1.069 & & $0.53\pm 0.4$ &$1 ^{h3}P_2 \rightarrow \gamma \gamma$ & 0.058 \\
$2 ^3P_2 \rightarrow \gamma \gamma$ & 0.7149 & 1.084& & &$2 ^{h3}P_2 \rightarrow \gamma \gamma$ & 0.1304\\
$3 ^3P_2 \rightarrow \gamma \gamma$ & 0.8686 & 1.0846& & &$3 ^{h3}P_2 \rightarrow \gamma \gamma$ & 0.2037\\
$4 ^3P_2 \rightarrow \gamma \gamma$ & 0.9956 &1.080 & & &$4 ^{h3}P_2 \rightarrow \gamma \gamma$ & 0.2758\\
$5 ^3P_2 \rightarrow \gamma \gamma$ & 1.1076 &1.077 & & &$5 ^{h3}P_2 \rightarrow \gamma \gamma$ & 0.3460 \\
$6 ^3P_2 \rightarrow \gamma \gamma$ & 1.209& & & &$6 ^{h3}P_2 \rightarrow \gamma \gamma$ & 0.4141\\ \hline

\end{tabular}
\end{center}
\end{table}

\begin{table}\caption{Three photon decay for $^3 S_1$ state conventional $c\overline{c}$ and hybrid mesons in $keV$. }
\begin{center}
\begin{tabular}{|c|c|c|c|} \hline
Decay Modes & Calculated $\Gamma_{\gamma\gamma\gamma}$ & Hybrid decay mode & Calculated $\Gamma_{\gamma\gamma\gamma}$\\ \hline
$1^3S_1 \rightarrow \gamma \gamma \gamma$ & 11.3992 &$1^{3h}S_1 \rightarrow \gamma \gamma \gamma$ & $1.0055 \times 10^{-3}$  \\
$2 ^3S_1 \rightarrow \gamma \gamma \gamma$ &5.275 &$2 ^{3h}S_1 \rightarrow \gamma \gamma \gamma$ &$1.2833 \times 10^{-3}$ \\
$3 ^3S_1 \rightarrow \gamma \gamma \gamma$ &4.1788 & $3 ^{3h}S_1 \rightarrow \gamma \gamma \gamma$ &$1.4162 \times 10^{-3}$  \\
$4 ^3S_1 \rightarrow \gamma \gamma \gamma$ &3.6871 & $4 ^{3h}S_1 \rightarrow \gamma \gamma \gamma$ &$1.4943 \times 10^{-3}$ \\
$5 ^3S_1 \rightarrow \gamma \gamma \gamma$ & 3.3968& $5 ^{3h}S_1 \rightarrow \gamma \gamma \gamma$ & $1.5449 \times 10^{-3}$ \\
$6 ^3S_1 \rightarrow \gamma \gamma \gamma$ & 3.2013& $6 ^{3h}S_1 \rightarrow \gamma \gamma \gamma$ & $1.5795 \times 10^{-3}$ \\ \hline
\end{tabular}
\end{center}
\end{table}

\section*{IV. Discussion and conclusion}

   The calculated masses reported in Tables (1,2) show that hybrids are more massive than corresponding conventional meson. As evident from decay constants, $|R(0)|2$, $|R(0)|^2$ reported in Tables(3-6), radial wave function at origin and decay constants for S states decreases toward higher radial excitations, while $|R'(0)|^2 $ and decay constants for P states increases toward higher radial excitations. It is also observed that decay constants for conventional meson are greater value as compared to the corresponding hybrid meson state. Results show that pseudo scalar $c\overline{c}$ mesons have higher values of $|R(0)|^2$ and $f_p$ as compare to vector mesons., Leptonic decay widths are reported for conventional and hybrid charmonium meson in Tables (7-8), while two photon and three photon decay widths for conventional and hybrid mesons are reported in Table (9-10). Results show that lepton and photon decay widths for hybrid mesons are more smaller than the conventional mesons. Calculated mass and radial wave function at origin and decay constants are in good agreement with theexperimental results \cite{PDG}. Calculated lepton and photon decay widths are close to experimental findings. In some cases, these results are more closer to model calculated results as compared to the experimental results\cite{PDG}.

\section{Acknowledgement}
The author acknowledges the financial support of Higher Education Commission of Pakistan through NRPU project number 7969.

\end{document}